\documentclass[aps,epsfig]{revtex4}%
\usepackage{graphicx}
\usepackage{epsfig}
\usepackage{amssymb}
\usepackage{bm}
\usepackage{amsmath}
\usepackage{amsfonts}%

\begin{document}
{\hfill Preprint JINR-E2-86-847,
Dubna, 1986}

\title{Effects of QCD Vacuum and Stability of $H$ Dihyperon
\footnote{Misprints are corrected, literature is updated.
Reprinted with partial support from RFBR grant no. 04-02-16445.}}
\author{A. E. Dorokhov, N. I. Kochelev\footnote{High Energy Physics Institute, AN
Kazakh. SSR}}
\affiliation{Joint Institute for Nuclear Research, Laboratory of Theoretical Physics,
140980, Dubna, Moscow Region, Russia}

\begin{abstract}
Within the composite quark model taking into account the interaction of quarks
in the bag with vacuum fields of QCD the masses of $H$ $(I=0,S=-2,J=0,1^{C})$
and $H^{\ast}$ $(I=0,S=-2,J=1,1^{C})$ dihyperons are estimated: $M_{H}=2.16$
GeV, $M_{H^{\ast}}=2.34$ GeV. It is shown that the leading effect giving a
stable with respect to strong decays $H$ dihyperon is the instanton
interaction forming diquarks: $q^{2}(J=0,C=\underline{\overline{3}})$,
$q^{2}(J=0,C=\underline{6})$. With the approach developed the contribution of
QCD hyperfine interaction is suppressed, and instanton induced three-particle
forces in multiquark hadrons ($q^{3}(J=1/2,C=\underline{8})$ - channel) are rather small.

\end{abstract}
\maketitle


1. Within the quark bag model
\cite{Bogolubov:1968zk,Chodos:1974je,Thomas:1982kv} the consideration of
multiquark states
\cite{Jaffe:1976ih,Jaffe:1976yi,Matveev:1977xt,Dorkin:1981ey} was one of the
most interesting applications. In these calculations the MIT version of bag
model was used. As proposed in a static approximation \cite{Bogolubov:1968zk},
the energy of a multiquark system is determined
\cite{Chodos:1974je,Donoghue:1979ax} by%
\begin{equation}
E\left(  R\right)  =\sum_{flav}\frac{n_{i}\omega_{i}}{R}+\frac{4\pi}{3}%
BR^{3}-\frac{Z_{0}}{R}+\Delta E_{g}, \label{1}%
\end{equation}%
\begin{equation}
M^{2}\left(  R\right)  =E^{2}-\left\langle P^{2}\right\rangle ,\qquad
\left\langle P^{2}\right\rangle \simeq\sum_{flav}n_{i}\left(  \frac
{\varkappa_{i}}{R}\right)  ^{2}, \label{2}%
\end{equation}
where $R$ is the bag radius, $n_{i}$ is the number of quarks of an $i$-th type
with energy $\omega_{i}/R$ ($\omega_{u,d}=2.043$ in the ground $s$-state;
$m_{u}=m_{d}=0$, $m_{s}\cong280$ MeV), $B$ is an external pressure
($B^{1/4}\cong145$ MeV), $-Z_{0}/R$ is a contribution of zero-mode
fluctuations ($Z_{0}=1.84$), $\Delta E_{g}$represents a color-magnetic
interaction ($\alpha_{s}=2.2$). The stability condition $\frac{dM^{2}}{dR}=0$
fixes the bag radius.

In the MIT version of the model the hadron spectrum is specified by the QCD
interaction%
\begin{equation}
\Delta E_{g}=-\frac{\alpha_{s}}{4R}\sum_{i>j}^{N}\mu\left(  m_{i}%
R,m_{j}R\right)  \left(  \overrightarrow{\sigma}\lambda^{a}\right)
_{i}\left(  \overrightarrow{\sigma}\lambda^{a}\right)  _{j}, \label{3}%
\end{equation}
where $N$ is the total number of quarks, $\overrightarrow{\sigma}_{i}\left(
\lambda_{i}^{a}\right)  $ are the spin (color) operators of an $i$-th quark;
$\mu_{ij}\equiv\mu\left(  m_{i}R,m_{j}R\right)  $ determine the strength of a
color-spin interaction. In a massless ($m_{i}=0$) case the average of
(\ref{3}) over a hadron state $q^{n}$ is expressed through the Casimir
operators of spin $SU_{2}^{J}$: $\frac{4}{3}J(J+1)$; color $SU_{3}^{C}$:
$C_{3}=0$; spin-color $SU_{6}^{CJ}$ - defined by quantum numbers of the
considered state%
\begin{equation}
\Delta E_{g}=\mu_{00}\left[  8N-\frac{1}{2}C_{6}+\frac{4}{3}J(J+1)\right]
\frac{\alpha_{s}}{4R}, \label{4}%
\end{equation}
where $J$ is the total moment. Based on this formula it is possible to
formulate the rules \cite{Jaffe:1976ih} analogous to the Hund rules from
atomic physics.

For the case of dibaryons rules are the following \cite{Jaffe:1976yi}: The
lightest states are those in which quarks are in the most symmetrical
(antisymmetrical) with respect to color-spin (flavor) representations.

Such general considerations and calculations based on the MIT model lead to
the conclusion \cite{Jaffe:1976yi} that a flavor-singlet six-quark dihyperon
$H$ with strangeness $-2$ and $J^{P}=0^{+}$ may be stable with respect to a
strong decay.

Recently, the problem of stability of the $H$ dihyperon has again been
discussed after unusual signals from the Cygnus X-3 have been registered
\cite{Marshak:1985ks}. As a possible explanation of this effect it was
proposed \cite{Baym:1985tn} that Cygnus X-3 is a star containing the strange
matter and it emits $H$ dibaryons with a lifetime $\tau_{H}\gtrsim10$ years.
The value of $H$ mass is an essential point in the determination of its
lifetime and confirmation of this hypothesis \cite{Khriplovich:1985wu}. While
today experimental situation on the emission from Cygnus X-3 stays indefinite
\cite{Berger:1986hn}, the search for multiquark states in cosmic rays and on
accelerators intensively continues \cite{DubnaVIII1986}. That is why correct
calculations of multiquark masses, their lifetimes and decay modes are so
important. In this respect the $H$ dihyperon is most intriguing object of the investigations.

The mass of $H$ was estimated within different variants of the bag model
\cite{Jaffe:1976yi,Matveev:1977xt,Liu:1982wg}, the lattice approach, the
QCD\ sum rules method and Skyrme model \cite{Mackenzie:1985vv}. The aim of the
present work is to calculate the mass of $H$ in the quark bag model taking
into account the structure of QCD vacuum \cite{Dorokhov:1986au}. In
\cite{Dorokhov:1986au} such a model was proved to be consistent with the
method of QCD sum rules and capable of describing the spectroscopy of the
ground states of hadrons.

2. Let us formulate the basic assumptions of the model \cite{Dorokhov:1986au}.
It is known \cite{Chodos:1974je} that in the MIT model, it is assumed
conserning the vacuum structure that in the presence of valence quarks
nonperturbative vacuum fully goes out of the bag. However, this hypothesis is
not compatible with the picture produced by the QCD sum rules
\cite{Chetyrkin:1978ta}, and it is not self-consistent. Really, the bag
constant $B$ characterizing the degree of destroying of vacuum
$B_{\mathrm{MIT}}\sim\left(  130-150\quad\mathrm{MeV}\right)  ^{4}$
\cite{DeGrand:1975cf} is much smaller than the "depth" of nonperturbative
vacuum known from the QCD sum rules: $\varepsilon_{0}=-\frac{9}{32}%
\left\langle 0\left\vert \frac{\alpha_{s}}{\pi}G^{2}\right\vert 0\right\rangle
\simeq-\left(  240\quad\mathrm{MeV}\right)  ^{4}$. This means that the vacuum
fields practically do not change inside the bag-hadron. Therefore, the neglect
of the effects of QCD vacuum in the bag becomes physically groundless.

In works \cite{Dorokhov:1986au} basic principles were formulated allowing one
to consider the QCD vacuum in the bag.

As a starting point, we take the QCD theory and available information on the
behavior of its solution. First, we think that in a zero approximation the
structure of the solution is such that the low- and high-frequency components
of the solution are independent of each other.

Second, we suppose that the (valence) components of the fields with
characteristic frequencies $\omega\sim\omega_{q}$ are described by solutions
of the static bag-model equations $\left(  q\left(  x\right)  =q^{\mathrm{bag}%
}\left(  x\right)  ;\quad A_{\mathrm{T}}\left(  x\right)  =A_{\mathrm{T}%
}^{\mathrm{bag}}\left(  x\right)  \right)  $.

Third, low-frequency (condensate) field components ($\omega<<\omega_{q}$) are
assumed to be solutions of the QCD equations characterized by a set of
numbers: different vacuum condensate quantities ($\left\langle \overline
{Q}Q\right\rangle $, $\left\langle G^{2}\right\rangle $,...). Under these
assumptions the Hamiltonian of the interaction of valence components
($q\left(  x\right)  ,A_{\text{\textrm{T}}}\left(  x\right)  $) with
condensate ones ($Q\left(  x\right)  ,A_{\text{\textrm{vac}}}\left(  x\right)
$) is restored uniquely through the field transformation:%
\begin{equation}
\Psi\left(  x\right)  =q\left(  x\right)  +Q\left(  x\right)  ;\qquad A\left(
x\right)  =A_{\mathrm{T}}\left(  x\right)  +A_{\mathrm{vac}}\left(  x\right)
. \label{5}%
\end{equation}
In addition, there are QCD vacuum fluctuations with $\omega_{\mathrm{vac}%
}>>\omega_{q}$ which may be approximated by the 't Hooft interaction
\cite{'tHooft:1976fv} induced by instantons.

3. In the model \cite{Dorokhov:1986au} the energy of a hadron is defined as%
\begin{equation}
M^{2}=E^{2}-\left\langle P^{2}\right\rangle ,\qquad\frac{dM^{2}}{dR}=0,
\label{6}%
\end{equation}
where $\left\langle P^{2}\right\rangle \simeq\sum_{\mathrm{flav}}n_{i}\left(
\varkappa_{i}/R\right)  ^{2}$ is due to the c.m. motion of quarks
\cite{Donoghue:1979ax} and%
\begin{equation}
E\left(  R\right)  =E_{\mathrm{kin}}+\Delta E_{\mathrm{g}}+\Delta
E_{\mathrm{vac}}+\Delta E_{\mathrm{inst}} \label{7}%
\end{equation}
is the bag energy. In (\ref{7}) the kinetic energy of quarks $E_{\mathrm{kin}%
}$ and the one-gluon interaction energy $\Delta E_{\mathrm{g}}$ are calculated
as usually in the bag perturbation theory
\cite{Bogolubov:1968zk,DeGrand:1975cf,Myhrer:1980jy}:%
\begin{equation}
E_{\mathrm{kin}}=\sum_{\mathrm{flav}}n_{i}\frac{\omega_{i}}{R}, \label{8}%
\end{equation}%
\begin{equation}
\Delta E_{\mathrm{g}}=\frac{0.117\alpha_{s}}{R}\left[  M_{00}+\left(
1-0.13m_{s}R\right)  M_{0s}+\left(  1-0.25m_{s}R\right)  M_{ss}\right]  ,
\label{9}%
\end{equation}
where $M_{ij}$ denotes matrix elements of the operator (\ref{2}) with respect
to spin-color spin states of hadrons.

As was shown in \cite{Dorokhov:1986au} a leading contribution to the hadron
energy caused by the valence- and condensate-fields interaction is generated
by the Hamiltonian%
\begin{equation}
H_{\mathrm{vac}}=\frac{\omega_{q}}{2}\left(  \overline{Q}\gamma^{0}%
q+\overline{q}\gamma^{0}Q\right)  . \label{10}%
\end{equation}
Then by using stationary perturbation theory%
\begin{equation}
\Delta E_{\mathrm{vac}}=\frac{\left\langle \Phi\left\vert H_{I}\right\vert
\Psi\right\rangle }{\left\langle \Phi|\Psi\right\rangle },\qquad\left\vert
\left.  \Psi\right\rangle \right.  =U\left(  -\infty,0\right)  \left\vert
\left.  \Phi\right\rangle \right.  , \label{11}%
\end{equation}%
\[
U\left(  -\infty,0\right)  =\sum_{n=0}^{\infty}\frac{\left(  -i\right)  ^{n}%
}{n!}\int_{-\infty}^{0}dt_{1}...\int_{-\infty}^{t_{n-1}}dt_{n}T\left[
H_{\mathrm{vac}}\left(  t_{1}\right)  ...H_{\mathrm{vac}}\left(  t_{n}\right)
\right]  ,
\]
where $\left\vert \left.  \Phi\right\rangle \right.  $ is a nonperturbed
hadron wave function of the bag model, we have \cite{Dorokhov:1986au}%
\begin{align}
\Delta E_{\mathrm{vac}}  &  =-n_{0}\frac{\pi}{24}\frac{\left\langle
0\left\vert \overline{u}u\right\vert 0\right\rangle }{\varkappa_{0}-1}%
R^{2}-n_{s}\frac{\pi}{12}\frac{\left\langle 0\left\vert \overline
{s}s\right\vert 0\right\rangle }{\varkappa_{s}^{2}}R^{2}\frac{\left(
y+a\right)  ^{2}y}{2y\left(  y-1\right)  +a}+\label{12}\\
&  +\frac{\pi^{2}}{1152}\frac{\left\langle 0\left\vert \overline
{u}u\right\vert 0\right\rangle ^{2}}{\varkappa_{0}\left(  \varkappa
_{0}-1\right)  ^{2}}R^{5}\left\{  \widetilde{M}_{00}+\frac{\left(
\varkappa_{0}+y\right)  \left(  y+a\right)  ^{2}\left(  \varkappa
_{0}-1\right)  }{\varkappa_{s}^{2}\left[  2y\left(  y-1\right)  +a\right]
}\frac{\left\langle 0\left\vert \overline{s}s\right\vert 0\right\rangle
}{\left\langle 0\left\vert \overline{u}u\right\vert 0\right\rangle }%
\widetilde{M}_{s0}+\right. \nonumber\\
&  +\left.  \frac{4y\left(  y+a\right)  ^{4}\varkappa_{0}\left(  \varkappa
_{0}-1\right)  ^{2}}{\varkappa_{s}^{4}\left[  2y\left(  y-1\right)  +a\right]
^{2}}\left(  \frac{\left\langle 0\left\vert \overline{s}s\right\vert
0\right\rangle }{\left\langle 0\left\vert \overline{u}u\right\vert
0\right\rangle }\right)  ^{2}\widetilde{M}_{ss}\right\}  +...,\nonumber
\end{align}
where $\left\langle 0\left\vert \overline{Q}_{i}Q_{i}\right\vert
0\right\rangle $ are quark condensates, $y=\omega_{s}R$, $a=m_{s}R$.

Expressions (\ref{12}) are absolutely different from $BR^{3}$ arising
\textit{ed hoc} in the MIT version. In contrast, in the model considered
stability of the bag is achieved in a self-consistent manner due to the
interaction of quarks with a physical vacuum. Moreover, the potential
(\ref{12}) is drastically dependent on the number of quarks with a given
flavor, their masses and quantum numbers of considered hadron states (the
latter is taken into account by the coefficients $\widetilde{M}$
\cite{Dorokhov:1986au}).

So, the long-wave vacuum fluctuations define the effective quark mass
(\ref{12}). At the same time the interaction of quarks with a short-wave part
of vacuum fluctuations allows us, to a great extent, to explain the mass
splitting between the terms of $SU_{f}(3)$ hadron multiplets
\cite{Dorokhov:1986au}. Within the model of QCD vacuum as an instanton liquid
\cite{Shuryak:1983nr} we get the two-particle contribution to the energy
\cite{Dorokhov:1986au}%
\begin{equation}
\Delta E_{\mathrm{inst}}^{(2)}=-n_{0}\sum_{\substack{a>b\\\mathrm{flav}}%
}\eta_{ab}I_{ab}\left\{  1+\frac{3}{32}\lambda_{a}\lambda_{b}\left(
1+3\overrightarrow{\sigma}_{a}\overrightarrow{\sigma}_{b}\right)  \right\}
\label{13}%
\end{equation}
and the contribution of three-particle forces for multiquark states%
\begin{align}
\Delta E_{\mathrm{inst}}^{(3)} &  =+n_{0}\eta_{uds}I_{uds}\left\{  1+\frac
{3}{32}\lambda_{a}\lambda_{b}\left(  1+3\overrightarrow{\sigma}_{a}%
\overrightarrow{\sigma}_{b}\right)  +\right.  \label{14}\\
&  -\frac{9}{320}d^{\alpha\beta\gamma}\lambda_{u}^{\alpha}\lambda_{d}^{\beta
}\lambda_{s}^{\gamma}\left[  1-3\left(  \overrightarrow{\sigma}_{a}%
\overrightarrow{\sigma}_{b}+\mathrm{permutations}\right)  \right]
+\nonumber\\
&  \left.  -\frac{9}{64}f^{\alpha\beta\gamma}\varepsilon_{ijk}\left(
\sigma^{i}\lambda^{\alpha}\right)  _{u}\left(  \sigma^{j}\lambda^{\beta
}\right)  _{d}\left(  \sigma^{k}\lambda^{\gamma}\right)  _{s}\right\}
.\nonumber
\end{align}
Here $n_{0}=\left\langle 0\left\vert g^{2}G^{a\mu\nu}G_{\mu\nu}^{a}\right\vert
0\right\rangle /\left(  64\pi^{2}\right)  $ is a density of instantons in the
model \cite{Shuryak:1983nr} $\left(  n_{0}=\left(  \pi\rho_{c}\left\langle
\overline{Q}Q\right\rangle \right)  ^{2}/3\right)  $,%
\[
\eta_{i_{1}...i_{n}}=\left(  \frac{4}{3}\pi^{2}\rho_{c}^{3}\right)
^{n}/\left(  m_{i_{1}}^{\ast}\rho_{c}...m_{i_{n}}^{\ast}\rho_{c}\right)  ,
\]%
\[
m_{i}^{\ast}=m_{i}+m^{\ast},\qquad m^{\ast}=-\frac{2\pi^{2}}{3}\left\langle
0\left\vert \overline{u}u\right\vert 0\right\rangle \rho_{c}^{2},
\]
$\rho_{c}$ is the characteristic size of an instanton in the QCD vacuum,%
\[
I_{a_{i}...a_{n}}=\int_{\mathrm{bag}}d\overrightarrow{r}%
{\displaystyle\prod\limits_{i=1}^{n}}
\overline{q}_{R}^{a_{i}}q_{L}^{a_{i}}.
\]

It should be emphasized that the interaction through instantons (\ref{13}),
(\ref{14}) takes place in a system $\left\vert {}\right\rangle $ of quarks in
a zero mode
\cite{'tHooft:1976fv,Shifman:1979uw,Betman:1985jj,Horn:1978cf,Dorokhov:1986au}%
\begin{equation}
\sum_{i=1}^{n}\left(  \overrightarrow{\sigma}_{S}^{i}+\overrightarrow{\tau
}_{C}^{i}\right)  \left\vert {}\right\rangle =0. \label{15}%
\end{equation}
Diquarks: $q^{2}\left(  \underline{\overline{3}}^{F},J=0,\overline{3}%
^{C}\right)  $, $q^{2}\left(  \underline{\overline{3}}^{F},J=1,6^{C}\right)
$, triquarks: $q^{3}\left(  \underline{8}^{F},J=\frac{1}{2},8^{C}\right)  $,
$q^{3}\left(  J=\frac{3}{2},10^{C}\right)  $, \textit{etc.} are such systems.

We also note that the instanton interaction (\ref{13}), (\ref{14}) is
consistent only in the first order of perturbation theory analogously to that
as it was done in the case of an external pion field \cite{DeTar:1980rr}.

4. The model parameters.

As is seen from (\ref{1}), the MIT model uses four parameters $\left(
B,\text{ }\alpha_{s},\text{ }Z_{0},\text{ }m_{s}\right)  $. But, the parameter
$Z_{0}$ is not well grounded, the value of $m_{s}$ is too large, and $B$
poorly agrees with the parameters extracted from the QCD sum rules and current
algebra. The value $\alpha_{s}=2.2$ does not agree with a perturbative
expansion in this parameter, which was confirmed in one-loop calculations
\cite{Goldhaber:1986ih}. In multiquark systems, perturbative calculations with
large $\alpha_{s}$ get still more uncertain \cite{Matveev:1977xt}.

Within the description \cite{Dorokhov:1986au} of energies of the ground states
of hadrons it has been proved that it suffices to choose the following values
of the parameters%
\begin{equation}
\alpha_{s}=0.7,\qquad m_{s}=220\quad\mathrm{MeV,} \label{16}%
\end{equation}
and $\rho_{c}=2$ GeV$^{-1}$, $\left\langle 0\left\vert \overline{Q}_{i}%
Q_{i}\right\vert 0\right\rangle =-\left(  250\quad\mathrm{MeV}\right)
^{3}\quad\left(  i=u,d,s\right)  $ adopted from the model of vacuum
\cite{Shuryak:1983nr} and QCD sum rules \cite{Chetyrkin:1978ta}, respectively.

5. To calculate the matrix elements of two- and three-particle operators
included into $\Delta E_{g},$ $\Delta E_{\mathrm{vac}}$, $\Delta
E_{\mathrm{inst}}^{(2)}$, $\Delta E_{\mathrm{inst}}^{(3)}$ it is necessary to
know the cluster expansion (dissociation) of the six-quark wave function of
$H$: $q^{6}\rightarrow q^{3}\times q^{3}$, $q^{6}\rightarrow q^{4}\times
q^{2}$. The expansion method and wave functions are given in Appendix.

By using the wave functions (\ref{A3}) we have for the matrix elements $\Delta
E_{g}$, $\Delta E_{\mathrm{inst}}^{(2)}$:%
\begin{align}
\Delta E_{g}^{H} &  =\left(  -5\mu_{00}-22\mu_{0s}+3\mu_{ss}\right)
/R,\label{17}\\
\Delta E_{g}^{H^{\ast}} &  =\frac{1}{3}\left(  \frac{11}{6}\mu_{00}-41\mu
_{0s}+\frac{67}{6}\mu_{ss}\right)  /R,\nonumber
\end{align}%
\begin{align}
\Delta E_{\mathrm{inst}}^{(2)H} &  =-\frac{27}{4}n_{0}\left(  I_{ud}\eta
_{ud}+2I_{us}\eta_{us}\right)  ,\label{18}\\
\Delta E_{\mathrm{inst}}^{(2)H^{\ast}} &  =-\frac{43}{8}n_{0}\left(
I_{ud}\eta_{ud}+2I_{us}\eta_{us}\right)  .\nonumber
\end{align}
In accordance with the selection rule (\ref{15}) the three-particle
interaction is nonzero only for the component $q^{3}\left(
I=0,S=-1,J=1/2,\underline{70}^{CJ}\right)  $. So, with the wave function
(\ref{A3}) we have:%
\begin{equation}
\Delta E_{\mathrm{inst}}^{(3)H}=+\frac{135}{8}n_{0}I_{uds}\eta_{uds}%
.\label{19}%
\end{equation}

It should be added that the approximation of coefficients $\mu$ is given in
\cite{Myhrer:1980jy}; coefficients $I_{ij}$ in \cite{Dorokhov:1986au} $\left(
I_{s}\eta_{s}\simeq0.65I_{0}\eta_{0}\right)  ,$ $I_{uds}=0.0244/R^6$.

6. By using the above-mentioned relations we obtain the estimation of
dihyperon masses%
\begin{align}
M_{H} &  =2.16\quad\mathrm{GeV,\qquad}R_{H}=5.2\quad\mathrm{GeV}%
^{-1}\mathrm{,}\label{20}\\
M_{H^{\ast}} &  =2.34\quad\mathrm{GeV,\qquad}R_{H^{\ast}}=5.3\quad
\mathrm{GeV}^{-1}\mathrm{.}\nonumber
\end{align}

So, our results show that the mass of $H$ is less than $2M_{\Lambda}$ but
above the threshold of $N\Lambda$. Dihyperon $H^{\ast}$ is absolutely
unstable: $M_{H^{\ast}}>2M_{\Lambda}$.

In accordance with the estimation of the lifetime of $H$ in $\Delta T=1$ weak
decays established in work \cite{Donoghue:1986zd}, the state with the mass
$M_{H}=2.16$ GeV is long-lived: $\tau_{H}\sim10^{-8}$ sec.

Note that in the approach developed a basic cause of the stability of $H$
dihyperon is the interaction of valence quarks with short-wave fluctuations
and physically is due to the same mechanism by which the mass splitting arises
in hadron multiplets ($\pi-\rho$, $N-\Delta$, and so on splittings).

We also proved the spectroscopic Hund rule for quark systems
\cite{Jaffe:1976ih,Jaffe:1976yi,Matveev:1977xt}. At the same time its origin
is absolutely different. The nonperturbative instanton interaction between a
pair of quarks produces strong attraction in a symmetric in color-spin
representation and is totally absent in antisymmetric states. The instanton
interaction takes into account the strong interaction at intermediate
distances $\rho_{c}$ and gives use to the formation of quasibound states,
diquarks \cite{Betman:1985jj,Fredriksson:1982qs}. So, the Hund rule is
physically due to the existence of diquarks.

Moreover, in multiquark system there are multiparticle $\left(  n>2\right)  $
instanton-induced forces (in colorless baryons such interactions are absent
because of the selection rule (\ref{15})). At the same time we show that the
contribution of three-particle forces to the energy of $H$ state is rather
small, $\Delta E_{\mathrm{inst}}^{(3)}\sim+70$ MeV.

Note also that very recently in the experiment carried by the B.A. Shahbasian
group the data that confirm the existence of $H$ dihyperon have been obtained
\cite{Shakhbazian:1986rf}.

\begin{acknowledgments}
In conclusion, the authors are thankful to P.N. Bogolubov for constant
attention and interest in the work and to S.B. Gerasimov, M.K. Volkov, A.V.
Radyushkin, A.I. Titov, S.M. Dorkin, V.T. Kim and B.A. Shahbasian for
discussions. One of us (N.K.) also thanks I.Ya. Chasnikov for support.
\end{acknowledgments}

\section*{Appendix. The wave function of dihyperons.}


\renewcommand{\theequation}{A.\arabic{equation}} \setcounter{equation}{0} To
construct the wave functions of $H$ and $H^{\ast}$ dihyperons, we make use of
the dissociation method developed in \cite{Bickerstaff:1980ji}. The idea of
this method was borrowed from work \cite{Matveev:1977xt}. That method allows
us to construct the wave functions of multiquark systems $q^{m}\overline
{q}^{n}$ with respect to arbitrary dissociation $\left(  q^{m_{1}}\overline
{q}^{n_{1}}\right)  \times\left(  q^{m_{2}}\overline{q}^{n_{2}}\right)  $
$\left(  m_{1}+m_{2}=m,n_{1}+n_{2}=n\right)  $. To classify the basis states
of multiquarks $\left(  q^{m}\right)  $, the group $U_{18}$ is chosen as the
group in which the direct production of isospin $SU_{2}^{I}$, strangeness
$U_{1}^{S}$, spin $SU_{2}^{J}$ and color $SU_{3}^{C}$ groups is embedded.
(Singling out the flavor group $SU_{3}^{F}$ is not effective because of a
strong mixing of the $SU_{3}^{F}$ quantum numbers arising in the multiquark
sector.) Thus, the quark wave function relative to their quantum numbers
transforms by a fundamental representation $\left\{  1\right\}  $ of the group
$U_{18}$. In this case the scheme of the group reduction of the $n$%
-quark-system representation $\left\{  1^{n}\right\}  $ is the following%
\begin{align}
U_{18}^{\times n}  &  \rightarrow U_{18}\rightarrow SU_{12}^{IJC}\times
U_{6}^{SJC}\rightarrow\label{A1}\\
&  \rightarrow\left(  SU_{2}^{I}\times SU_{6}^{JC}\right)  \times\left(
U_{1}^{S}\times SU_{6}^{JC}\right)  \rightarrow\nonumber\\
&  \rightarrow SU_{2}^{I}\times\left(  SU_{2}^{J}\times SU_{3}^{C}\right)
\times U_{1}^{S}\times\left(  SU_{2}^{J}\times SU_{3}^{C}\right)
\rightarrow\nonumber\\
&  \rightarrow SU_{2}^{I}\times U_{1}^{S}\times SU_{2}^{J}\times SU_{3}%
^{C}.\nonumber
\end{align}

Using Racah's factorization lemma (1949) and the factorization property of
transformation coefficients for direct product groups, the transformation of a
$6$-quark wave function to the dissociation basis may be written in the form
\cite{Bickerstaff:1980ji}%
\begin{equation}
\left\vert \left(  1^{3},1^{3}\right)  1^{6}\left(  1^{n_{0}}\left(
I,\lambda_{0}^{JC},J_{0}\mu_{0}^{C}\right)  ,1^{n_{S}}\left(  S,\lambda
_{S}^{JC},J_{S}\mu_{S}^{C}\right)  \right)  J0^{C}\right\rangle = \label{A2}%
\end{equation}%
\begin{align*}
&  =\sum_{\left(  ^{\prime},^{\prime\prime}\right)  }\left\vert \left[
1^{3}\left(  1^{n_{0}^{\prime}}\left(  I^{\prime},\lambda_{0}^{\prime}%
,J_{0}^{\prime}\mu_{0}^{\prime}\right)  ,1^{n_{S}^{\prime}}\left(  S^{\prime
},\lambda_{S}^{\prime},J_{S}^{\prime}\mu_{S}^{\prime}\right)  \right)
J^{\prime}\mu^{\prime};\right.  \right. \\
&  \left.  \left.  1^{3}\left(  1^{n_{0}^{\prime\prime}}\left(  I^{\prime
\prime},\lambda_{0}^{\prime\prime},J_{0}^{\prime\prime}\mu_{0}^{\prime\prime
}\right)  ,1^{n_{S}^{\prime\prime}}\left(  S^{\prime\prime},\lambda
_{S}^{\prime\prime},J_{S}^{\prime\prime}\mu_{S}^{\prime\prime}\right)
\right)  J^{\prime\prime}\mu^{\prime\prime}\right]  ISJ0^{C}\right\rangle \\
&  \left\langle \left(  1^{3}\left(  1^{n_{0}^{\prime}}1^{n_{S}^{\prime}%
}\right)  ;1^{3}\left(  1^{n_{0}^{\prime\prime}}1^{n_{S}^{\prime\prime}%
}\right)  \right)  |\left(  1^{3},1^{3}\right)  1^{6}\underline{\left(
1^{n_{0}}1^{n_{S}}\right)  }\right\rangle \\
&  \left\langle \left(  1^{n_{0}^{\prime}}\left(  I^{\prime}\lambda
_{0}^{\prime}\right)  ;1^{n_{0}^{\prime\prime}}\left(  I^{\prime\prime}%
\lambda_{0}^{\prime\prime}\right)  \right)  |\left(  1^{n_{0}^{\prime}%
},1^{n_{0}^{\prime\prime}}\right)  1^{n_{0}}\underline{\left(  I\lambda
_{0}\right)  }\right\rangle \\
&  \left\langle \left(  1^{n_{S}^{\prime}}\left(  S^{\prime}\lambda
_{S}^{\prime}\right)  ;1^{n_{S}^{\prime\prime}}\left(  S^{\prime\prime}%
\lambda_{S}^{\prime\prime}\right)  \right)  |\left(  1^{n_{S}^{\prime}%
},1^{n_{S}^{\prime\prime}}\right)  1^{n_{S}}\underline{\left(  S\lambda
_{S}\right)  }\right\rangle \\
&  \left\langle \left(  \lambda_{0}^{\prime}J_{0}^{\prime}\mu_{0}^{\prime
};\lambda_{0}^{\prime\prime}J_{0}^{\prime\prime}\mu_{0}^{\prime\prime}\right)
|\left(  \lambda_{0}^{\prime}\lambda_{0}^{\prime\prime}\right)  \lambda
_{0}\underline{\left(  J_{0}\mu_{0}\right)  }\right\rangle \\
&  \left\langle \left(  \lambda_{S}^{\prime}J_{S}^{\prime}\mu_{S}^{\prime
};\lambda_{S}^{\prime\prime}J_{S}^{\prime\prime}\mu_{S}^{\prime\prime}\right)
|\left(  \lambda_{S}^{\prime}\lambda_{S}^{\prime\prime}\right)  \lambda
_{S}\underline{\left(  J_{S}\mu_{S}\right)  }\right\rangle \\
&  \left\langle \left(  J_{0}^{\prime}J_{S}^{\prime}\right)  J^{\prime
},\left(  J_{0}^{\prime\prime}J_{S}^{\prime\prime}\right)  J^{\prime\prime
}|\left(  J_{0}^{\prime}J_{0}^{\prime\prime}\right)  J_{0},\left(
J_{S}^{\prime}J_{S}^{\prime\prime}\right)  J_{S}\left(  \underline{J}\right)
\right\rangle \\
&  \left\langle \left(  \mu_{0}^{\prime}\mu_{S}^{\prime}\right)  \mu^{\prime
},\left(  \mu_{0}^{\prime\prime}\mu_{S}^{\prime\prime}\right)  \mu
^{\prime\prime}|\left(  \mu_{0}^{\prime}\mu_{0}^{\prime\prime}\right)  \mu
_{0},\left(  \mu_{S}^{\prime}\mu_{S}^{\prime\prime}\right)  \mu_{S}\left(
\underline{0^{C}}\right)  \right\rangle .
\end{align*}
Here $n_{0}\left(  n_{S}\right)  $ is the number of $\left(  u,d\right)  $
(and $s$) quarks, $I,S$ and $J$ are the isospin, strangeness and total spin,
respectively, the representation of $SU_{6}^{JC}\left(  SU_{3}^{C}\right)  $.

The complete expression of the expansion of the $H$ dihyperon wave function
(the $H^{\ast}$ wave function was given in ref. \cite{Bickerstaff:1980ji}) is:%
\begin{equation}
\left\vert H\right\rangle =0.867\left\vert e_{1}\right\rangle -0.499\left\vert
e_{2}\right\rangle , \label{A3}%
\end{equation}%
\begin{align*}
\left\vert e_{1}\right\rangle  &  =\sqrt{\frac{2}{15}}\frac{1}{\sqrt{2}%
}\left\{  Q_{0}\left(  \frac{1}{2},21,^{4}21\right)  Q_{-2}\left(  1^{2}%
,^{3}1^{2};^{4}21\right)  -\longleftrightarrow\right\}  -\\
&  -\sqrt{\frac{2}{15}}\frac{1}{\sqrt{2}}\left\{  Q_{0}\left(  \frac{1}%
{2},21,^{2}21\right)  Q_{-2}\left(  1^{2},^{3}1^{2};^{2}21\right)
-\longleftrightarrow\right\}  +\\
&  +\sqrt{\frac{2}{15}}\frac{1}{\sqrt{2}}\left\{  Q_{0}^{N}\left(  \frac{1}%
{2},21,^{2}0\right)  Q_{-2}^{\Xi}\left(  1^{2},^{3}1^{2};^{2}0\right)
-\longleftrightarrow\right\}  -\\
&  -\sqrt{\frac{1}{10}}\frac{1}{\sqrt{2}}\left\{  Q_{-1}\left(  1,1^{2}%
,^{1}2;^{2}21\right)  Q_{-1}\left(  1,1^{2},^{3}1^{2};^{2}21\right)
+\longleftrightarrow\right\}  +\\
&  +\sqrt{\frac{2}{45}}\left\{  Q_{-1}^{\Sigma}\left(  1,1^{2},^{3}1^{2}%
;^{2}0\right)  \right\}  ^{2}-\sqrt{\frac{1}{45}}\left\{  Q_{-1}^{\Sigma
^{\ast}}\left(  1,1^{2},^{3}1^{2};^{4}0\right)  \right\}  ^{2}+\\
&  +\sqrt{\frac{2}{45}}\left\{  Q_{-1}\left(  1,1^{2},^{3}1^{2};^{4}21\right)
\right\}  ^{2}-\sqrt{\frac{4}{45}}\left\{  Q_{-1}\left(  1,1^{2},^{3}%
1^{2};^{2}21\right)  \right\}  ^{2}-\\
&  -\sqrt{\frac{3}{10}}\frac{1}{\sqrt{2}}\left\{  Q_{-1}\left(  0,2,^{1}%
1^{2};^{2}21\right)  Q_{-1}\left(  0,2,^{3}2;^{2}21\right)
+\longleftrightarrow\right\}  ,
\end{align*}%
\begin{align*}
\left\vert e_{2}\right\rangle  &  =\sqrt{\frac{2}{5}}\frac{1}{\sqrt{2}%
}\left\{  Q_{0}\left(  \frac{1}{2},21,^{2}21\right)  Q_{-2}\left(  1^{2}%
,^{1}2;^{2}21\right)  -\longleftrightarrow\right\}  +\\
&  +\sqrt{\frac{3}{20}}\left\{  Q_{-1}\left(  1,1^{2},^{1}2;^{2}21\right)
\right\}  ^{2}-\sqrt{\frac{1}{30}}\left\{  Q_{-1}^{\Sigma}\left(  1,1^{2}%
,^{3}1^{2};^{2}0\right)  \right\}  ^{2}-\\
&  -\sqrt{\frac{1}{15}}\left\{  Q_{-1}^{\Sigma^{\ast}}\left(  1,1^{2}%
,^{3}1^{2};^{4}0\right)  \right\}  ^{2}-\sqrt{\frac{1}{60}}\left\{
Q_{-1}\left(  1,1^{2},^{3}1^{2};^{2}21\right)  \right\}  ^{2}-\\
&  -\sqrt{\frac{1}{30}}\left\{  Q_{-1}\left(  1,1^{2},^{3}1^{2};^{4}21\right)
\right\}  ^{2}+\sqrt{\frac{1}{10}}\left\{  Q_{-1}^{\Lambda}\left(
0,2,^{1}1^{2};^{2}0\right)  \right\}  ^{2}+\\
&  +\sqrt{\frac{1}{20}}\left\{  Q_{-1}\left(  0,2,^{1}1^{2};^{2}21\right)
\right\}  ^{2}+\sqrt{\frac{1}{20}}\left\{  Q_{-1}\left(  0,2,^{3}%
2;^{2}21\right)  \right\}  ^{2}+\\
&  +\sqrt{\frac{1}{10}}\left\{  Q_{-1}\left(  0,2,^{3}2;^{4}21\right)
\right\}  ^{2},
\end{align*}
where%
\begin{align*}
Q_{0}\left(  \frac{1}{2},21,^{2}21\right)   &  =Q^{3}\left(  n_{0}%
=3;I_{0}=\frac{1}{2},\lambda_{0}=21,J_{0}=\frac{1}{2},\mu_{0}=21\right)  ,\\
Q_{-1}\left(  0,2,^{3}2;^{2}21\right)   &  =Q^{3}\left(  n_{0}=2;I_{0}%
=0,\lambda_{0}=2,J_{0}=1,\mu_{0}=2;J=\frac{1}{2},\mu=21\right)  ,\\
Q_{-2}\left(  1^{2},^{3}1^{2};^{4}21\right)   &  =Q^{3}\left(  n_{0}%
=1;\lambda_{S}=1^{2},J_{S}=1,\mu_{S}=1^{2};J=\frac{3}{2},\mu=21\right)  .
\end{align*}

Expression (\ref{A3}) and the analogous one for $H^{\ast}$ from
\cite{Bickerstaff:1980ji} are used to calculate matrix elements of the
three-particle operator contained in $\Delta E_{\mathrm{inst}}^{\left(
3\right)  }$. The scalar isospin components with the strangeness $-1:$
$Q_{-1}\left(  0,2,^{1}1^{2};^{2}21\right)  $ and $Q_{-1}\left(
0,2,^{3}2;^{2}21\right)  $ only give a nonzero contribution. These components
are weighted with the probability $25\%$ in the total wave function.

To compute averages of two-particle operators, one may apply expressions of
kind (\ref{A2}) or to construct the dissociation $q^{6}\rightarrow q^{4}\times
q^{2}$. The coefficients of this dissociation may be easily found by using the
equations for the Casimir operators of the $SU_{2}^{J}$, $SU_{3}^{F}$ and
$SU_{3}^{C}$ groups. As a result, we have for the basis $SU_{3}^{F}\times
SU_{2}^{J}\times SU_{3}^{C}$%
\begin{align}
\left\vert H\left(  0^{F},0^{J},0^{C}\right)  \right\rangle  &  =\sqrt
{\frac{1}{10}}q^{4}\left(  2,0,2\right)  q^{2}\left(  2,0,2\right)
+\sqrt{\frac{3}{10}}\left[  q^{4}\left(  1^{2},0,1^{2}\right)  q^{2}\left(
1^{2},0,1^{2}\right)  +\right. \label{A4}\\
&  \left.  +q^{4}\left(  1^{2},1,2\right)  q^{2}\left(  1^{2},1,2\right)
+q^{4}\left(  2,1,1^{2}\right)  q^{2}\left(  2,1,1^{2}\right)  \right]
,\nonumber
\end{align}%
\begin{align}
\left\vert H^{\ast}\left(  21^{F},1^{J},0^{C}\right)  \right\rangle  &
=\sqrt{\frac{7}{60}}q^{4}\left(  2,0,2\right)  q^{2}\left(  2,0,2\right)
+\sqrt{\frac{13}{60}}q^{4}\left(  1^{2},0,1^{2}\right)  q^{2}\left(
1^{2},0,1^{2}\right)  +\label{A5}\\
&  +\sqrt{\frac{17}{60}}q^{4}\left(  1^{2},1,2\right)  q^{2}\left(
1^{2},1,2\right)  +\sqrt{\frac{23}{60}}q^{4}\left(  2,1,1^{2}\right)
q^{2}\left(  2,1,1^{2}\right)  .\nonumber
\end{align}
If the dissociations (\ref{A3})-(\ref{A5}) are expressed as
\begin{equation}
Q^{6}=\sum_{i}w_{i}\left(  Q^{3}\right)  _{i}^{\prime}\times\left(
Q^{3}\right)  _{i}=\sum_{j}u_{j}\left(  Q^{4}\right)  _{j}\times\left(
Q^{2}\right)  _{j}, \label{A6}%
\end{equation}
then the matrix elements of the three-$\left(  R_{3}\right)  $ and
two-$\left(  R_{2}\right)  $ particle operators are calculated with the help%
\begin{equation}
\left\langle Q^{6}\left\vert R_{3}\right\vert Q^{6}\right\rangle =20\sum
w_{i}^{2}\left\langle Q_{i}^{3}\left\vert R_{3}\right\vert Q_{i}%
^{3}\right\rangle , \label{A7}%
\end{equation}%
\begin{equation}
\left\langle Q^{6}\left\vert R_{2}\right\vert Q^{6}\right\rangle =15\sum
u_{j}^{2}\left\langle Q_{j}^{2}\left\vert R_{2}\right\vert Q_{j}%
^{2}\right\rangle ,
\end{equation}
where the combinatorial factors take into account the multiple character of
interaction in a $n$-particle antisymmetric states.

\end{document}